\documentclass[amsmath,preprintnumbers,showpacs,twocolumn,plb,superscriptaddress, nofootinbib]{revtex4}
\usepackage{amssymb}
\usepackage{graphicx}
\usepackage{epstopdf}
\usepackage{bm}
\usepackage{epsfig}
\usepackage{graphics}
\usepackage{xspace}
\usepackage{amsfonts}
\usepackage{verbatim}
\usepackage{natbib}
\usepackage{color}

\begin{document}

\title{ Inverting the mass hierarchy of jet quenching effects with prompt $b$-jet substructure}
\author{Hai Tao Li} \email{haitaoli@lanl.gov}
\affiliation{
Los Alamos National Laboratory, Theoretical Division, Los Alamos, NM 87545, USA}
\author{Ivan Vitev} \email{ivitev@lanl.gov }
\affiliation{
Los Alamos National Laboratory, Theoretical Division, Los Alamos, NM 87545, USA}

\begin{abstract}
The mass of heavy quarks, such as charm and bottom, plays an important role in the formation of parton showers. This effect is apparently not well understood  when parton showers evolve in a strongly interacting quark-gluon plasma. We propose a new experimental measurement in relativistic heavy ion collisions,  based on a two-prong subjet structure  inside a reconstructed heavy flavor jet,  which can place stringent constraints on the mass dependence of  in-medium splitting functions. We identify the  region of jet transverse  momenta where parton mass effects are leading and predict a unique reversal of the mass hierarchy of jet quenching effects in heavy ion  relative to proton collisions. Namely, the momentum sharing distribution of prompt $b$-tagged jets is more strongly modified  in comparison to the one for light jets.  Our work is useful in guiding experimental efforts at the Large Hadron Collider and the Relativistic Heavy Ion Collider in the near future. 
\end{abstract}

\maketitle

\emph{ \bf Introduction}. Understanding the production and  structure of hadronic jets is crucial to test perturbative Quantum Chromodynamics (QCD) and make full use of 
the data from today's high energy collider experiments. With the ever increasing center-of-mass energies $\sqrt{s_{\rm NN}}$ of hadronic and heavy ion collisions, heavy quarks, such as charm ($c$) and bottom ($b$), are copiously produced in parton showers. The fraction of jets  initiated by prompt heavy quarks is also becoming sizable, necessitating more precise theoretical control on the effects of parton mass.  It was suggested more than a decade ago that these mass effects should be readily observable in heavy ion collisions, leading to reduced radiative energy losses of charm and bottom quarks relative to light  quarks~\cite{Dokshitzer:2001zm,Djordjevic:2003zk,Zhang:2003wk} in hot and dense nuclear matter.     

Experimental measurements of open heavy flavor and $b$-quark jets have not clearly established this ``dead cone" effect~\cite{Dokshitzer:2001zm}, which was predicted to result in smaller cross section
attenuation of $D$-mesons and, especially, $B$-mesons relative to light hadrons in ultrarelativistic nuclear collisions. At high transverse momenta the difference in the magnitude of heavy and light 
quark flavor quenching disappears due to the non-Abelian analog of the Landau-Pomeranchuk-Migdal (LPM) effect~\cite{Wang:1994fx} and the different hardness 
of light parton  and  heavy quark fragmentation functions. Recently, it was also pointed out that both open heavy flavor and $b$-jet production receive 
large contribution from gluon fragmentation into heavy flavor~\cite{Kang:2016ofv,Huang:2013vaa}.  At low, but perturbatively accessible, transverse momenta at the Relativistic Heavy Ion Collider (RHIC)  it was noticed early on that the  anticipated mass effect on $B$-meson quenching is not reflected in the suppression of 
non-photonic electrons coming from the semileptonic decays of open heavy flavor~\cite{Djordjevic:2005db}.  This discrepancy has stimulated  extensive theoretical work, for a comprehensive review that covers theory and experimental measurements see~\cite{Andronic:2015wma}, suggesting that collisional energy loss effects may play 
a very important role in this kinematic domain.  Only very recently has there been an indication at the Large Hadron Collider (LHC) that the suppression of $B$-mesons, inferred via the
$B\rightarrow J/\psi$ channel, might be smaller than that of $D$-mesons and light hadrons~\cite{Khachatryan:2016ypw}.    
 
It is, therefore, both timely and critical to identify new experimental observables, which are sensitive to the mass effects in parton branching and 
shower evolution.  Jet substructure~\cite{Larkoski:2017jix} is a promising direction to investigate,  and a recently proposed  study of the two leading  subjets inside a reconstructed jet~\cite{Larkoski:2015lea}  can accurately test the $1\to2$  QCD splitting function~\cite{Larkoski:2017bvj}.  The technique itself is based  on the ``soft drop grooming"~\cite{Larkoski:2014wba}, which removes soft wide-angle radiation from a jet until hard 2-prong substructure is found. The jet momentum sharing variable is then defined as 
\begin{align}
    z_g=\frac{\min(p_{T1}, p_{T2})}{p_{T1}+p_{T2}}~, \quad  z_g>z_{\rm cut} \left(\frac{\Delta R_{12}}{R} \right)^\beta \;, 
    \label{zg}
\end{align}
where $p_{T1}$ and $p_{T2}$ are the transverse momenta of the subjets. Soft bremsstrahlung is eliminated through the minimum  $z_g$ requirement, where  in Eq.~(\ref{zg}) $\Delta R_{12}$ is the distance between two subjets and $R$ is the radius of the original jet. In the  limit of large  jet energies the  distribution of $z_g$  maps directly onto the 
widely used lowest order Dokshitzer-Gribov-Lipatov-Altarelli-Parisi splitting functions.   

In heavy ion (A+A) collisions the interactions of the outgoing partons with the hot and dense QCD medium, the quark-gluon plasma (QGP), may change the jet splitting functions
relative to the simpler proton-proton (p+p) case. Pioneering experimental studies of this observable have recently been carried out  by the CMS collaboration~\cite{Sirunyan:2017bsd} at the LHC and the STAR collaboration~\cite{Kauder:2017cvz} at RHIC. Theoretically, it was shown that the nuclear modification of the jet momentum sharing distribution~\cite{Chien:2016led}  is directly related to the in-medium splitting functions~\cite{Ovanesyan:2011kn}, which can be obtained in the 
framework of soft-collinear effective theory with Glauber gluon interactions (SCET$_{\rm G}$)~\cite{Idilbi:2008vm,Ovanesyan:2011xy}.  Similar conclusion was reached in  other works focused on the
soft gluon emission  parton energy loss limit~\cite{Mehtar-Tani:2016aco,Chang:2017gkt}. Monte Carlo studies do not include the full LPM effect physics, but nevertheless show that naive  subjet energy loss does not lead to jet substructure modification~\cite{Lapidus:2017dek}.  A single model attributes the experimentally observed changes to non-perturbative effects that are not well understood at present~\cite{Milhano:2017nzm}.
Thus, the overwhelming preponderance of studies show that the momentum sharing distribution of jets not  only complements the extensive suite of jet quenching measurements in 
A+A reactions at RHIC~\cite{Timilsina:2016sjv,Adamczyk:2017yhe} and LHC~\cite{Aad:2010bu,Chatrchyan:2012nia,Aad:2012vca,Abelev:2013kqa,Chatrchyan:2013kwa,Chatrchyan:2014ava,Adam:2015ewa,Khachatryan:2015lha,Adam:2015doa,Aaboud:2017bzv,Acharya:2017goa,Sirunyan:2017qhf,Sirunyan:2017jic,ATLAS:2017isy}, but also provides a new handle on and direct  access to the fundamental many-body perturbative QCD splitting processes probed in heavy ion collisions.   \\

\emph{ \bf Theoretical formalism}. 
Recent advances in generalizing SCET$_{\rm G}$ to include finite heavy quark masses~\cite{Kang:2016ofv}  are the 
stepping stone for the first calculation of the heavy flavor jet splitting function in Au+Au collisions at RHIC and Pb+Pb collisions at the LHC, which we present in this Letter.  We are interested in the limit $0<m \ll p^+$~\cite{Rothstein:2003wh,Leibovich:2003jd}, where $m$ is the heavy quark mass and $p^+$ is the large lightcone momentum\footnote{For the case $m\sim p^+$, which is beyond the  scope of this Letter,  see~\cite{Fleming:2007qr}. }.  Consider an off-shell parton of momentum $[p^+ ,p^-,{\bf 0}_\perp]$  that
splits into two daughter partons $[zp^+ , {\bf k}^2_\perp/zp^+ ,{\bf k}_\perp]$ and $[(1-z)p^+ , {\bf k}^2_\perp/(1-z)p^+,- {\bf k}_\perp]$\footnote{${\bf k}_\perp,-{\bf k}_\perp$ are the subjet momenta  perpendicular to the parent parton direction, not to be confused with the transverse momentum of the jet in the lab frame.}. In the absence of a QCD medium 
 the massive vacuum splitting kernels $Q\to Qg$ , $Q\to gQ$, and $g\to Q\bar{Q}$ read
\begin{eqnarray} \label{eq:Msp1}
    \left(\frac{dN^{\rm vac}}{dz d^2 \mathbf{k}_\perp}\right)_{Q\to Qg} &=&  \frac{\alpha_s}{2 \pi^2} \frac{C_F}{\mathbf{k}_{\perp}^2+z^2m^2}  
   \\ & &
 \times   \left( 
    \frac{1+(1-z)^2}{z}-\frac{2z(1-z)m^2}{\mathbf{k}_{\perp}^2+z^2m^2}
    \right)~, \nonumber \\
    \label{eq:Msp2}
  \left(\frac{dN^{\rm vac}}{dz d^2 \mathbf{k}_\perp}\right)_{g\to Q\bar{Q}} &=&  \frac{\alpha_s}{2 \pi^2} \frac{ T_R}{\mathbf{k}_{\perp}^2+m^2} 
     \\ &&
   \times   \left( 
   z^2+(1-z)^2+\frac{2z(1-z)m^2}{\mathbf{k}_{\perp}^2+m^2}
    \right)~,    \nonumber \\ 
    \left(\frac{dN^{\rm vac}}{dz d^2 \mathbf{k}_\perp}\right)_{Q\to gQ} &=&       \left(\frac{dN^{\rm vac}}{dz d^2 \mathbf{k}_\perp}\right)_{Q\to Qg} (z\rightarrow 1-z) \; .
\label{eq:Msp3}
\end{eqnarray}
Here,  $C_F$ is the Casimir of the fundamental representation of SU(3) and $T_R=1/2$ is  the trace normalization of the fundamental representation.
 The above equations reduce to the massless splitting functions when $m=0$, and the $g \rightarrow gg$ kernel is well known and
not shown here. If we denote by $r_g \equiv \Delta R_{12}$ the angular separation between the two final state partons and $E_0=p^+/2$ is the energy,  
$\mathbf{k}_{\perp} = z(1-z) r_g E_0$ when $r_g$ is not too large.  To identify the region of phase space where heavy quark mass effects are leading,
we consider a typical separation between the two subjets inside a reconstructed jet $r_g=0.2$ and a momentum sharing 
fraction $z \sim 1/2$.  The essential  ${\bf k}_\perp^2 < z^2 m^2, m^2, (1-z)^2 m^2$ condition will be strictly satisfied for prompt $b$-jets of 
energy $E_0 \leq 25$~GeV, but non-trivial mass effects extend to energies at least twice as large. While those energies are smaller  than the jet energies first studied in heavy ion collisions at the LHC,  they are now accessible with improved  experimental jet  reconstruction techniques~\cite{Adamczyk:2017yhe}.  Such moderate energy jets  are also the cornerstone of the jet physics  program with the future sPHENIX experiment at  RHIC~\cite{sPHENIX:2015irh}. 
 
To set up the stage for the jet splitting function calculation in heavy ion collisions, we start with the vacuum case.   We denote by  $j\to i \bar{i}$  
the parton branchings and define $r_g=\theta_g R$. The $\theta_g$ and $z_g$ distribution for parton $j$,  after soft-drop grooming  is 
\begin{align}
    \left(\frac{dN^{\rm vac}}{dz_g d\theta_g}\right)_{j}  = \frac{\alpha_s}{\pi} \frac{1}{\theta_g} \sum_{i} P_{j\to i \bar{i}}^{\rm vac}(z_g)~.
\end{align}
When the splitting probability becomes large, resummation is necessary and was performed  to modified leading-logarithmic (MLL) accuracy
 in Ref.~\cite{Larkoski:2014wba}. The resummed distribution for a $j$-type jet, initiated by a massless quark or a gluon,  is 
\begin{align} \label{eq:mll}
    \frac{dN_j^{\rm vac,MLL}}{ dz_g d\theta_g} =& \sum_{i} \left(\frac{dN^{\rm vac}}{dz_g d\theta_g}\right)_{j\to i \bar{i}}
    \nonumber \\ &
     \underbrace{\exp \left[-\int_{\theta_g}^1 d\theta \int_{z_{\rm cut}}^{1/2} dz  \sum_{i} \left(\frac{dN^{\rm vac}}{dz d\theta}\right)_{j\to i \bar{i}}  \right]}_{\rm Sudakov~Factor}~.
\end{align}
The normalized joint probability distribution then reads
\begin{align}
    p(\theta_g,z_g)\big|_{j}=\frac{\frac{dN_j^{\rm vac,MLL}}{ dz_g d\theta_g} }{\int_{0}^1 d\theta \int_{z_{\rm cut}}^{1/2} dz  \frac{dN_j^{\rm vac,MLL}}{ dz d\theta} }~. 
\label{eq:jointprob}
\end{align}

\begin{figure*}[t]
    \centering
    \includegraphics[scale=0.8]{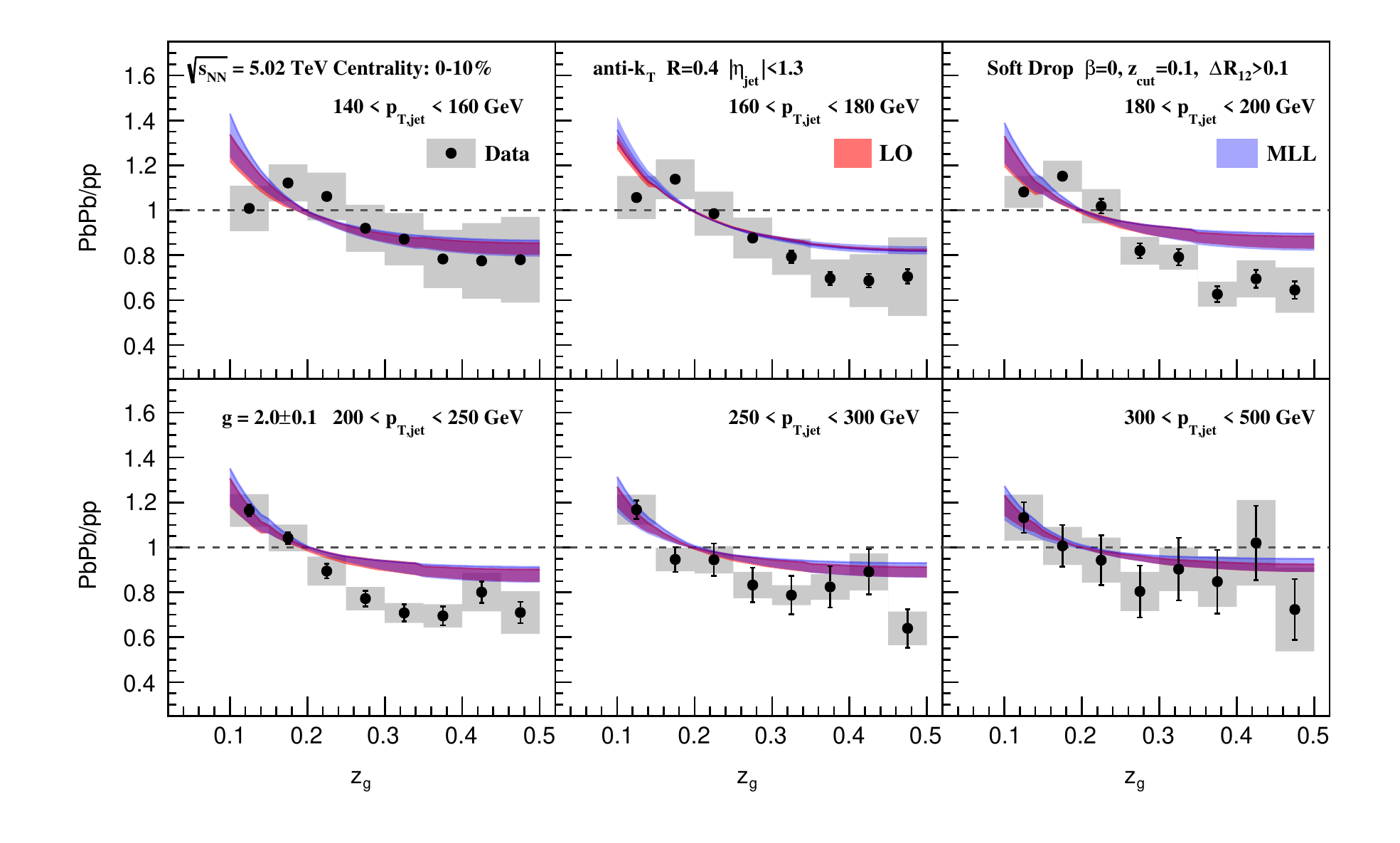}
    \vspace{-0.8cm}
    \caption{Comparison of theoretical predictions for the in-medium $z_g$ distribution modification to the CMS measurements~\cite{Sirunyan:2017bsd} in Pb+Pb collisions with different jet $p_T$ intervals at $\sqrt{s_{\rm NN}}=5.02$ TeV.}
    \label{fig:cms_all}
\end{figure*}

Suppose that we can distinguish the splitting process involving heavy flavor, for example by tagging jets and subjets with leading charm and beauty mesons ($D$, $B$).
In the absence of  a QCD medium,  such study was  proposed  in Ref.~\cite{Ilten:2017rbd} and simulations performed using a Monte Carlo event generator 
framework. Analytically, Eq.~(\ref{eq:jointprob}) can be extended to the 
case of a heavy flavor jet splitting, such as $b\to bg$ or $c\to cg $, in a straight forward way. For gluon splitting into heavy quark pairs the probability function is defined as 
\begin{align}
    p(\theta_g, z_g)\big|_{g\to Q\bar{Q}} = \frac{\left(\frac{dN^{\rm vac}}{dz_g d\theta_g}\right)_{g\to Q \bar{Q}} \Sigma_{g}(\theta_g)}{\int_{0}^{1}d\theta \int_{z_{\rm cut}}^{1/2} dz  \left(\frac{dN^{\rm vac}}{dz d\theta}\right)_{g\to Q \bar{Q}} \Sigma_{g}(\theta)  }~,
\end{align}
where $\Sigma_{g}(\theta_g)$ is the Sudakov factor for gluon evolution in Eq.~(\ref{eq:mll}) and it exponentiates all the possible contributions from gluon splitting, such as $g\to g g $ and $g\to q\bar{q}$. Thus, we find that MLL resummation can change significantly the predictions for the $g\to Q\bar{Q}$ channel relative to the leading order (LO) results.  
The final probability distribution for 
$z_g$ is defined as 
\begin{align}
    p(z_g)\big|_j &=  \frac{1}{\sigma_j}\int dp_T d \eta \frac{d\sigma_{j}}{dp_T d\eta } \int_{0}^{1}d\theta ~p(\theta, z_g)\big|_{j}~,    
\end{align}
where $\sigma_{j}$ is the cross section of the $j$-parton production. Strictly speaking, the quark or gluon production cross section is not well-defined. In the view of the perturbative nature of  higher order corrections, it is sufficient to use the LO predictions  of $\sigma_j $ for the current calculations~\cite{Tripathee:2017ybi}. We use MADGRAPH5\_AMC@NLO~\cite{Alwall:2014hca} and  
NNPDF2.3 LO PDF sets~\cite{Ball:2012cx} to generate the LO events for jet production.

In the presence of a QCD medium,  it was demonstrated that the vacuum splitting functions must be replaced by the full splitting kernels  for each possible channel 
\begin{align}
   \frac{dN^{\rm full}}{dz d^2 \mathbf{k}_\perp} = \frac{dN^{\rm vac}}{dz d^2 \mathbf{k}_\perp}   + \frac{dN^{\rm med}}{dz d^2 \mathbf{k}_\perp},        
\end{align}
where the complete sets of $dN^{\rm med}/dz d^2 \mathbf{k}_\perp$ for zero and finite quark masses on the right hand side  can be found in Refs.~\cite{Ovanesyan:2011kn,Kang:2016ofv}  to first order in opacity.  They have been applied  to describe and/or predict jet quenching effects for inclusive hadrons and heavy 
mesons~\cite{Kang:2014xsa,Chien:2015vja,Kang:2016ofv}, as well as jets and related jet substructure~\cite{Chien:2015hda,Chien:2016led,Kang:2017frl}, in fixed order and resummed 
calculations\footnote{Recently in-medium splitting kernels have been calculated to any order in opacity and  proof-of-principle numerical  evaluation shown to second order~\cite{Sievert:2019cwq}.}.  
In all perturbative calculations we rely on the large scale separation between the energetic particles and jets ${\cal O}(10-1000~{\rm GeV})$ and the  medium  ${\cal O}(100~{\rm MeV})$.
The full in-medium splitting kernels provide a systematic framework to study high transverse momentum  observables in the nuclear matter  environment  beyond the traditional energy loss approaches. 
We evaluate them in a QGP background simulated by  2+1-dimensional viscous event-by-event hydrodynamics~\cite{Shen:2014vra}, which was recently used to calculate
quarkonium suppression~\cite{Aronson:2017ymv} and, more importantly, $b$-jet suppression~\cite{Li:2018xuv} at the LHC.  \\

\begin{figure*}
    \centering
    \includegraphics[scale=0.43]{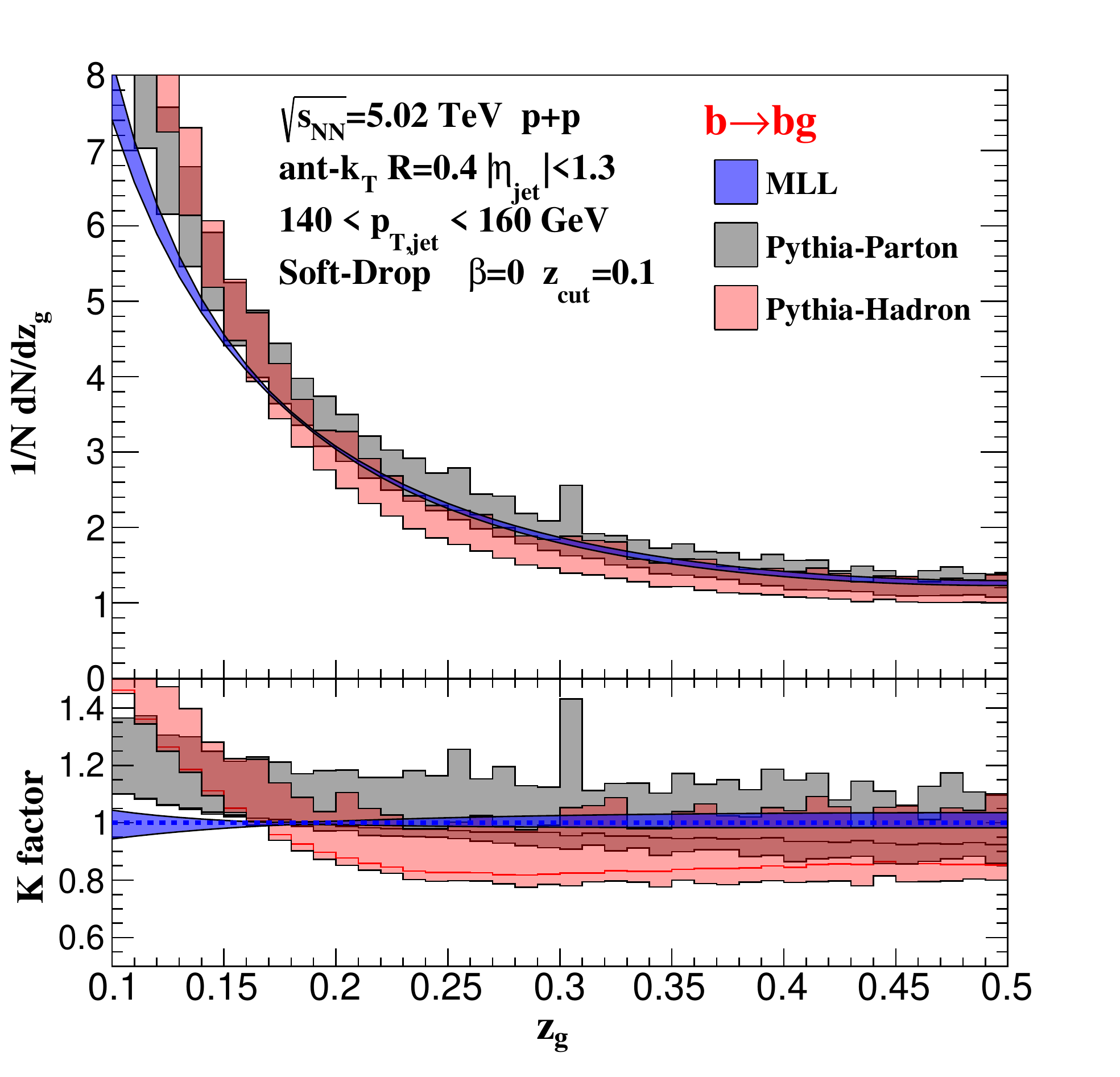}
    \includegraphics[scale=0.43]{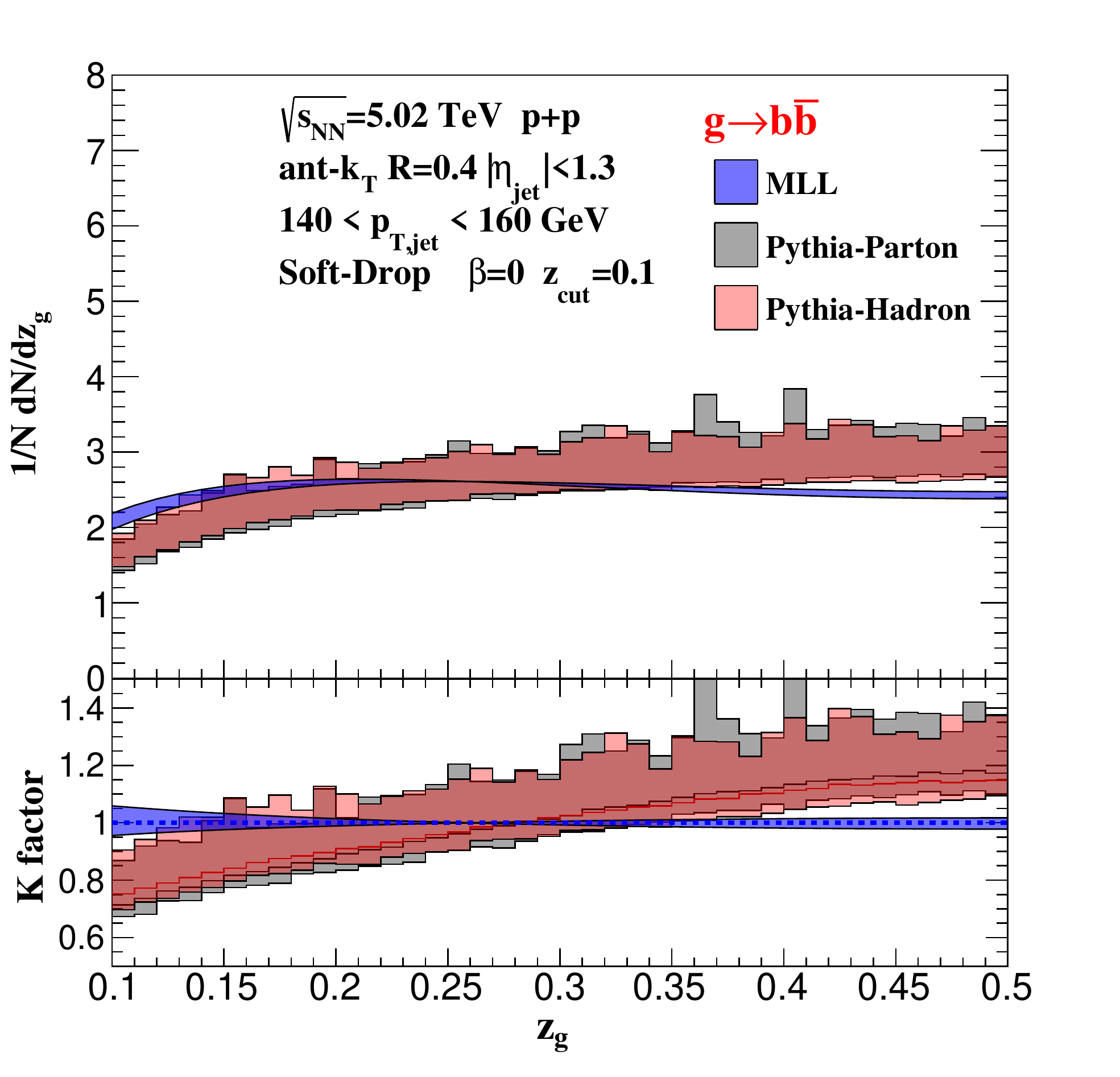}
        \vspace{-0.5cm}
    \caption{Momentum sharing distributions for heavy flavor tagged jet and the comparison with PYTHIA predictions in pp collisions. The gray and red bands represent the parton showers predictions with and without hadronization, respectively.  The blue band represents our MLL calculation. We present results for the  $b\rightarrow bg$ channel (left panel) and  $g\rightarrow b{\bar b}$ channel (right panel) for LHC energy $\sqrt{s_{\rm NN}}=5.02$~TeV.  }
    \label{fig:pythia}
\end{figure*}

\begin{figure*}
    \centering
    \includegraphics[scale=0.43]{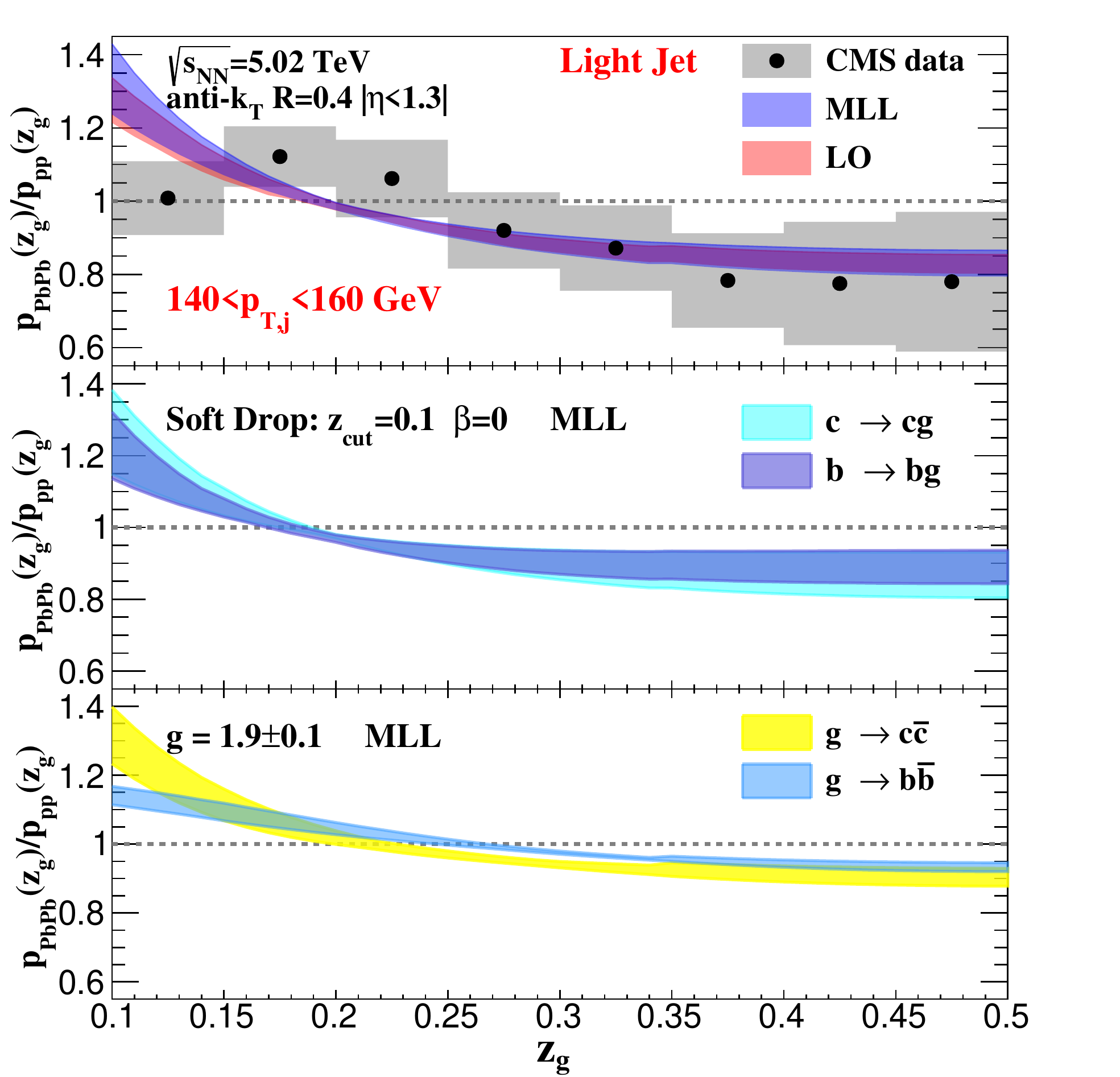}
     \includegraphics[scale=0.43]{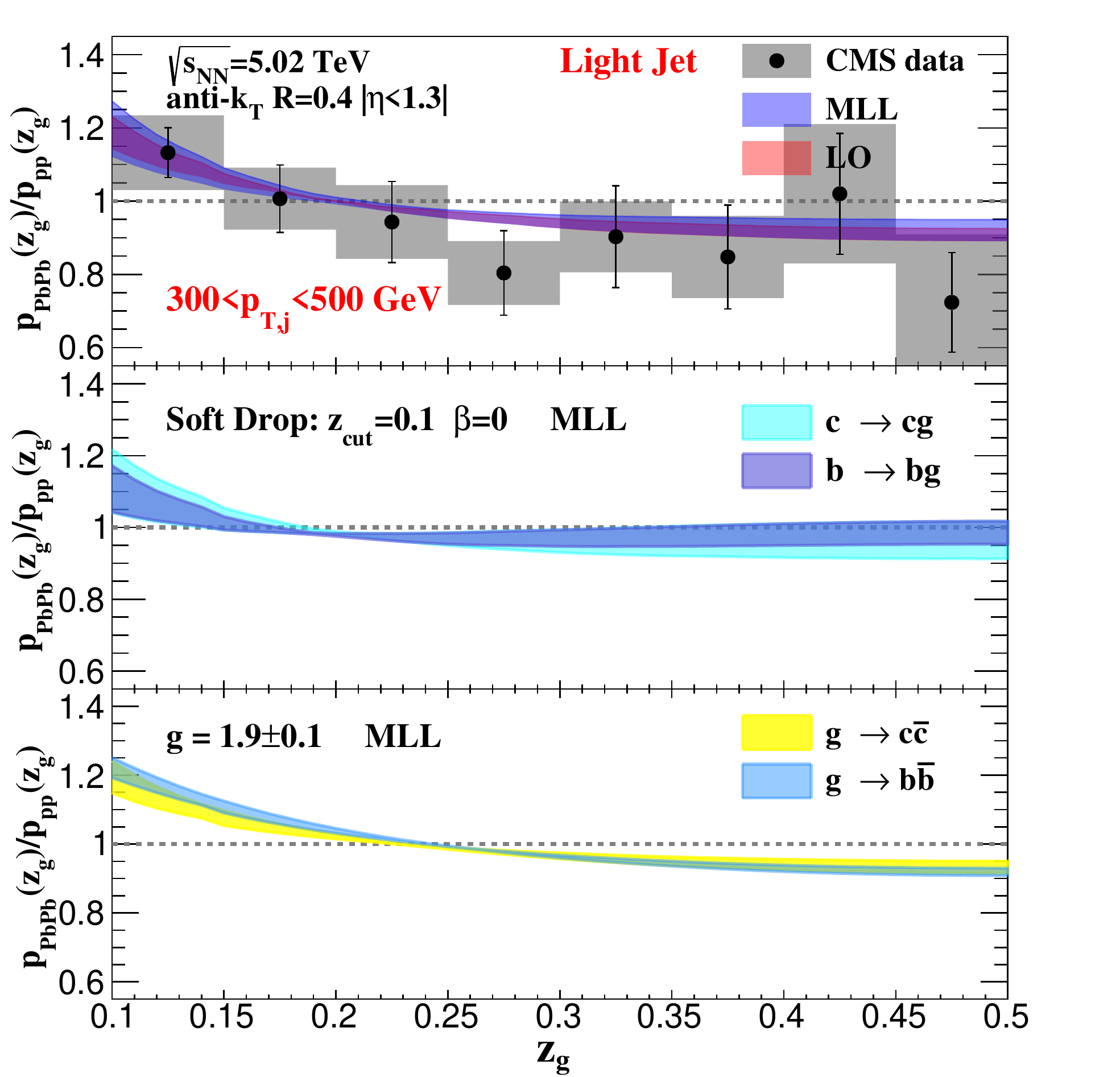}
    \vspace{-0.3cm}    
    \caption{The modification of the jet splitting functions  in 0-10\% central Pb+Pb  collisions at $\sqrt{s_{\rm NN}}=5.02$~TeV  for two $p_T$ bins  $140<p_{T,j}<160$~GeV (left panel) and $300<p_{T,j}<500$~GeV (right panel). The upper panels compare the LO and MLL predictions to CMS  light  jet substructure measurements. The middle  and lower panels 
    present the MLL modifications for heavy flavor tagged jet - the $Q\rightarrow Qg$ and $\rightarrow Q{\bar Q}$, respectively. }
    \label{fig:cms}
\end{figure*}

\emph{ \bf Numerical results}. 
 For all  predictions we use one-loop running  coupling $\alpha_s$ and choose the scale as $\max(\mu, \mu_{\rm NP})$,  $\mu_{\rm NP}$ being the non-perturbative value where we freeze the coupling. The default scale choices are $\mu=k_\perp$ and $\mu_{\rm NP} =1$~GeV. The renormalization and factorization scales for the LO jet production are chosen as $\mu_h = p_{T1}+p_{T2}$, where $p_{T1}$ and $p_{T2}$ are the transverse momenta of the jets. The uncertainties are obtained by varying $\mu$, $\mu_{\rm NP}$, $\mu_h$ by a factor of two independently.  We also include in all figures the uncertainty from the variation of the coupling $g$ between the jet and the QGP, which enters in the medium corrections of the splitting functions. The parameter settings in every figure of this Letter are the same for different panels, unless specified otherwise.

In Fig.~\ref{fig:cms_all} we first presents the modifications\footnote{The modifications are defined as the ratio of the $z_g$ distributions in the medium and the vacuum.} of the groomed light jet momentum sharing distributions, which are compared to the CMS measurements over different kinematic ranges in 0-10\% central Pb+Pb collisions  at $\sqrt{s_{\rm NN}}=5.02$~TeV~\cite{Sirunyan:2017bsd}. This observable has been studied before but the technical advance included in the new calculation is the resummation of the medium-induced radiation. Jets are reconstructed using anti-$k_T$ algorithm~\cite{Cacciari:2008gp} with $R=0.4$ and $|\eta|<1.3$ in both p+p and Pb+Pb collisions.  Besides to the jet $p_{T}$ and rapidity cut, an additional cut on the distance between the two subjets $\Delta R_{12}>0.1$ is applied due  to the detector resolution of the measurements, which makes it possible to provide LO predictions, as was done in~\cite{Chien:2016led}.  As a result of the cut, we notice in the upper panels that there is little difference  between the fixed order  and resummed  calculations.  Given the large uncertainties from the measurements, the jet quenching effects are described well by the medium induced splitting functions, perhaps with the exception of the large $z_g$  region at $p_T \sim 200$~GeV.

Before we proceed to the calculation of the $b$-jet momentum sharing distribution we will study the effect of non-perturbative hadronization corrections and compare the MLL  to Monte-Carlo  event generator simulations. Figure~\ref{fig:pythia} shows our predictions of the $z_g$ distributions for $b\to bg$ and $g\to b \bar{b}$ channels in pp collisions, as well as the predictions given by PYTHIA~\cite{Sjostrand:2007gs}. 
The parton shower generator uncertainties are obtained by varying the scale in the QCD strong coupling $\alpha_s$. The difference between PYTHIA predictions with and without hadronization give us the measure of non-perturbative effects. From   Fig.~\ref{fig:pythia}  we can conclude that the effects of hadronization on the $z_g$ distributions are small. They are slightly more pronounced in the $b\rightarrow bg$ channel in
comparison to the $g \rightarrow b{\bar b}$ channel, but even in this case the parton  level and hadron level simulations overlap within the theoretical uncertainties.  The MLL calculations are also consistent with the PYTHIA predictions, any remaining differences  are limited to
$\leq 20$\% and such differences typically cancel in the nuclear modification ratio.

For high $p_T$ jets in heavy ion collisions, where the mass effect is small,  the $b\to bg$ and $g\to b\bar{b}$ groomed $z_g$ distribution provides a way to study the  quark jet evolution in the medium and the effects of gluon splitting into heavy quark-antiquark pairs inside one  jet initiated by a gluon.  
In Fig.~\ref{fig:cms} we show  the modifications of the momentum sharing distributions  of  inclusive jets (in the upper panel for reference), the $Q\to Qg$  (middle panel) and $g\to Q\bar{Q}$  (bottom panel),   over the kinematic ranges  $140<p_{T, j}<160$ GeV and $300<p_{T, j}<400$ GeV in 0-10\% central Pb+Pb collisions  at $\sqrt{s_{\rm NN}}=5.02$~TeV.  We find that the predicted jet quenching effects for $p(z_g)$  of $Q\to Qg$ channels are comparable to that of light jets, and they can be measured by the LHC experiments. The modification of the $z_g$  distribution is somewhat smaller for $g\to b\bar{b}$ in comparison to the other splitting functions in the lower $p_{T,j}$ regions. By comparing the predictions for b-jet and c-jet modifications, we notice that the mass effect slowly vanishes with increasing jet energy.

\begin{figure}
    \centering
    \includegraphics[scale=0.45]{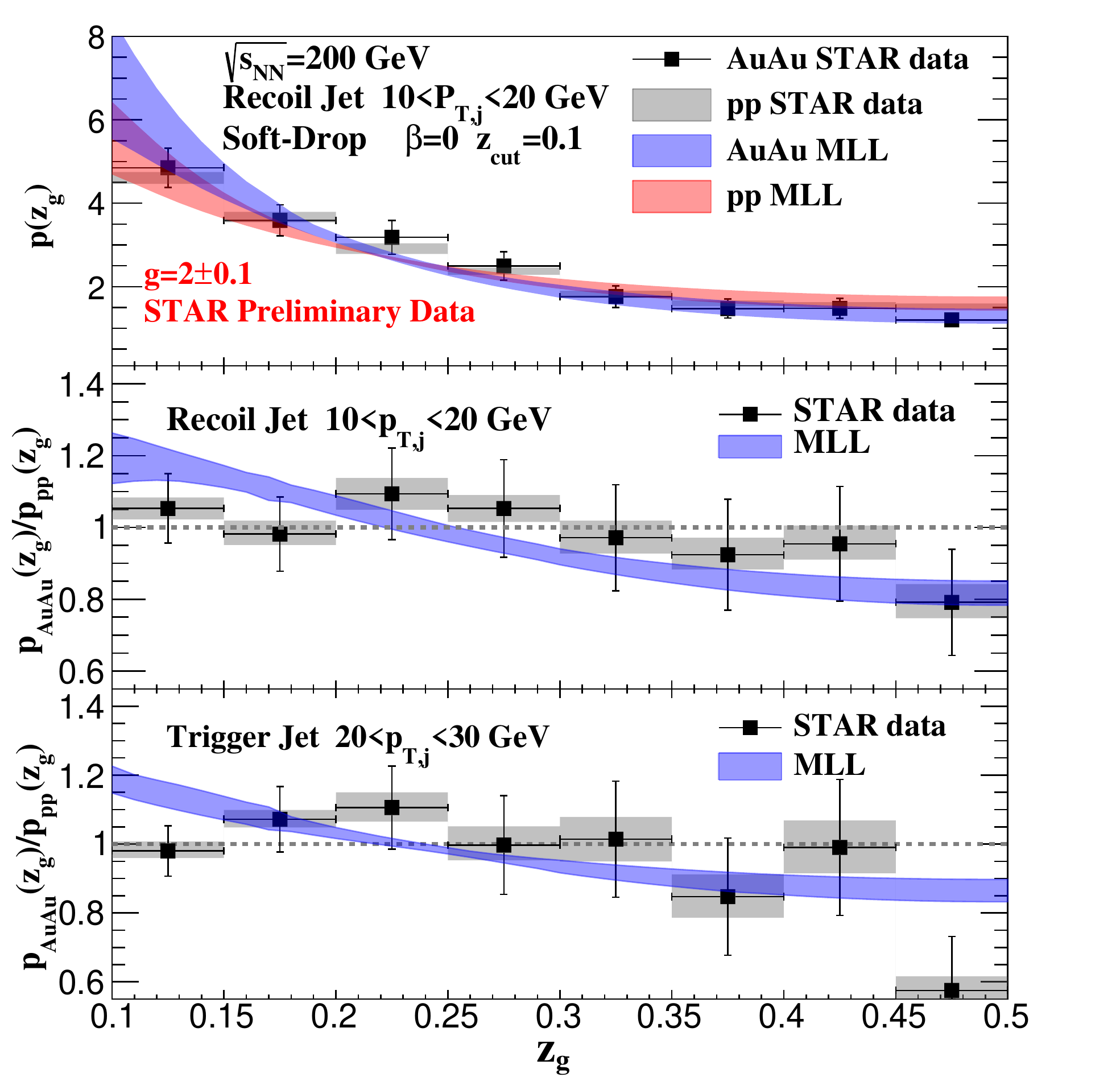}
    \vspace{-0.5cm}    
    \caption{ Distribution of $z_g$ and its modification  for recoil and trigger light jets at $\sqrt{s_{\rm NN}}=200$~GeV Au+Au collisions.  The upper panel compares the theoretical predictions for the recoil jet to the preliminary STAR data~\cite{Kauder:2017cvz}. The middle and bottom panels show the predictions and  measurements for the modification of recoil  and trigger jets, respectively.}
    \label{fig:star}
\end{figure}

Next, we turn to the jet momentum sharing distribution modification in Au+Au collisions at  RHIC.  The coupling $g$ is chosen to be slightly larger than the one for LHC because of the smaller collision energy at RHIC~\cite{Kang:2014xsa}.  Non-perturbative effects should be more important  at lower center-of-mass energies. However, we find that in the normalized $p(z_g)$ distribution and, especially, in the ratio $p_{\rm AuAu}(z_g)/p_{\rm pp}(z_g)$ the sensitivity to  non-perturbative physics is reduced. Before we address the question of heavy flavor jet substructure modification at RHIC, we turn to the case of light flavor jets.          
Figure~\ref{fig:star} compares the MLL jet splitting functions for the trigger and  recoil jets in p+p and Au+Au collisions at $\sqrt{s_{\rm NN}}=200$~GeV to measurements from the STAR collaboration~\cite{Kauder:2017cvz}. 
The shaded gray bands and the vertical error bars represent the experimental uncertainties in Au+Au collision and p+p collision, respectively. 
The theoretical calculations in heavy ion collisions take into account the geometric bias due to triggering and the slightly different 0-20\% centrality in comparison to the LHC results. An important difference between the CMS momentum sharing distribution measurement~\cite{Sirunyan:2017bsd} and the one done by STAR~\cite{Kauder:2017cvz} is that while the former uses a grooming radius $\Delta R_{12}>0.1$, the latter does not. This necessitates  the resummation of the  vacuum and in-medium branching processes, which we perform in this Letter.    
   Both of the splitting functions and the modification for the recoil jet are in good agreement with data. As we can see, both the MLL results and the measured modifications are somewhat smaller that those at the LHC. The bottom panel  of Fig.~\ref{fig:star} compares our calculation and the measurement for the trigger jet and they are consistent within experimental uncertainties.

\begin{figure}
    \centering
        \vspace{-0.5cm}
    \includegraphics[scale=0.45]{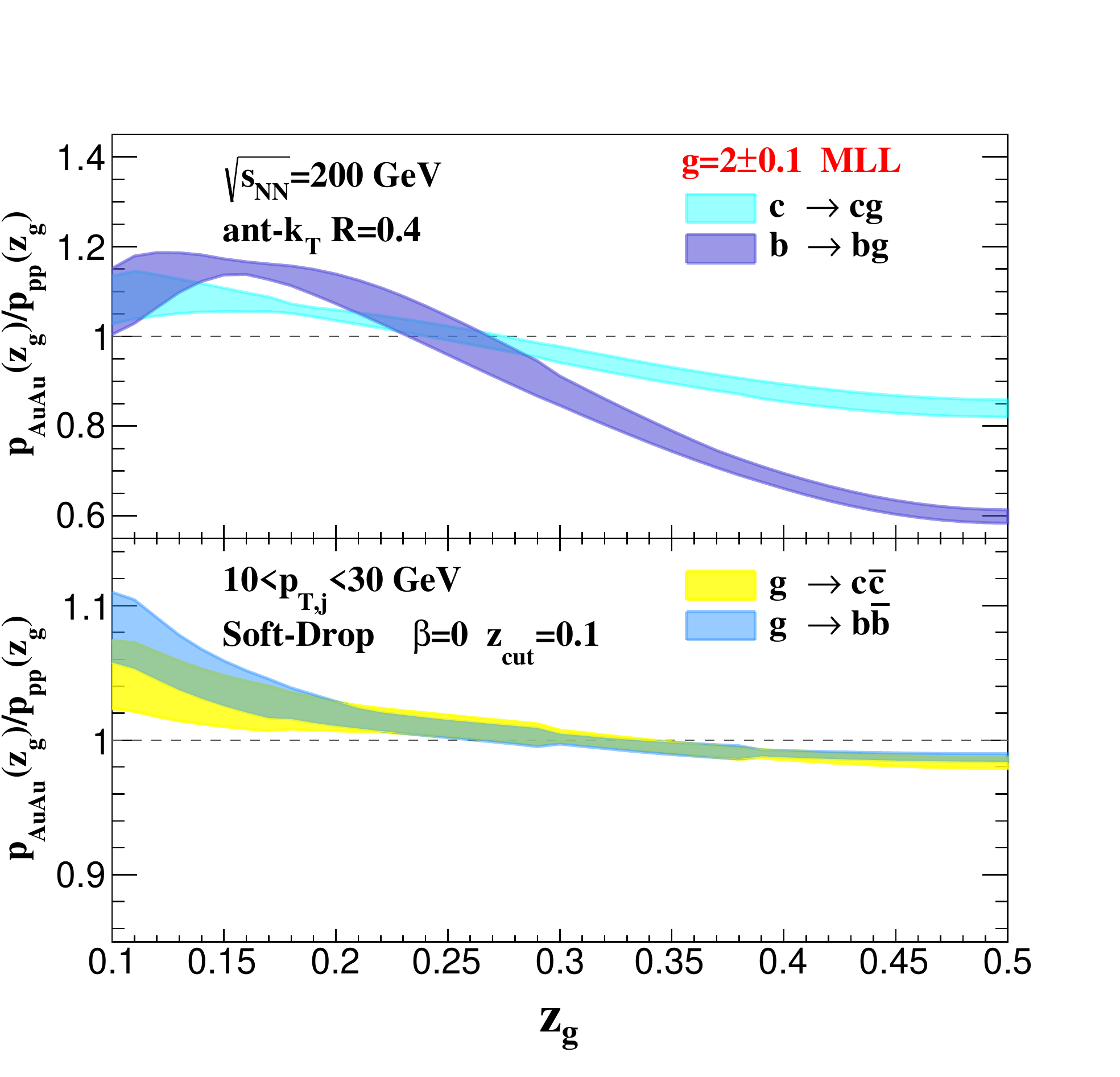}
    \vspace{-0.5cm}
    \caption{The modifications of the splitting functions for heavy flavor tagged jet at $\sqrt{s_{\rm NN}}=200$~GeV Au+Au collisions. Note the  strong quenching effects for prompt
    $b$-jets contrasted by the lack of any significant QGP-induced modification for the $g\to Q\bar{Q}$ splitting.}
    \label{fig:AuAu_Q}
\end{figure}

Before we turn to heavy flavor jets of lower transverse momenta, let us summarize the analytic first principles expectations for the jet momentum sharing distribution modification at fixed order.  We focus on the  way in which the mass terms in the denominator of the splitting kernels Eqs.~(\ref{eq:Msp1})-(\ref{eq:Msp3})  alter the longitudinal $z$ dependence of parton branching, noticing that in the medium one has two such mass dependent propagators~\cite{Kang:2016ofv}. If  $\mathbf{k}^2_\perp \ll z^2_g m^2$  the  $Q\rightarrow Qg$ distribution is considerably steeper that the one for light partons~\cite{Kang:2016ofv} amplifying the $p_{\rm AA}(z_g)$ versus the $p_{\rm pp}(z_g)$ difference.  Conversely, when the   $\mathbf{k}^2_\perp \ll  m^2$  the $z$ dependence in  the $g\to Q\bar{Q}$ channel is approximately constant (no $z$ dependence), leading to no nuclear modification. If 
we write the $z_g$ distribution in heavy ion collisions  $p_{AA}$ as normalized  $ (p_{pp} + p_{\rm med})$,  we can predict the following jet splitting function modifications
in order of decreasing strength
\begin{equation}
\frac{p_{med}^{Q\rightarrow Qg}(z_g)}{p_{\rm pp}^{Q\rightarrow Qg}(z_g)} \sim \frac{1}{z_g^2},   \; 
\frac{p_{med}^{j\rightarrow i\bar{i}}(z_g)}{p_{\rm pp}^{j\rightarrow i\bar{i}}(z_g)} \sim \frac{1}{z_g},   \; 
\frac{p_{med}^{g\rightarrow Q\bar{Q}}(z_g)}{p_{\rm pp}^{g\rightarrow Q\bar{Q}}(z_g)} \sim {\rm const} .
\label{order}
\end{equation}
Numerical results for the momentum sharing distribution ratios for heavy flavor tagged jets in  Au+Au to p+p collisions  at $\sqrt{s_{\rm NN}}=200$ GeV are presented in Fig.~\ref{fig:AuAu_Q}.  
We consider  $10<p_{T, j}<30$~GeV where our analysis suggests that heavy quark mass effects on parton shower formation are the largest, especially for bottom quarks.    For $c\to cg$, the  
$p(z_g)$ modification in the QGP is similar to the one for light jets, however, the $b\to bg$ channel  exhibits  much larger in-medium effects.  This  unique reversal in the mass hierarchy of jet quenching effects\footnote{We have checked that in the soft gluon emission limit parton energy loss ordering 
$\Delta E^{\rm rad}_b <  \Delta E^{\rm rad}_c < \Delta E^{\rm rad}_{u,d} < \Delta E^{\rm rad}_g$ holds. } 
is in perfect agreement  with Eq.~(\ref{order}), and so is the lack of any significant nuclear modification for  the $g\to Q\bar{Q}$ channel seen in the bottom panel of
Fig.~\ref{fig:AuAu_Q}.   We have also shown that the larger modification of the prompt $b$-jet splitting function persists to $p_T$ of more than 50~GeV at the LHC and can be measured there as well. \\

\emph{ \bf Conclusions}.  
 In this Letter, we presented the first resummed calculation of the soft-drop groomed momentum sharing distributions in heavy ion collisions in the framework of  recently developed effective theories of light parton and heavy quark  propagation in the dense QCD matter.  For light jets, the modification of this observable in Au+Au and Pb+Pb reactions agrees well with the recent experimental measurements over a wide range of center-of-mass energies, validating the theoretical approach. The most important advances reported in this work, however, relate to heavy flavor tagged jets. We demonstrated that jet splitting functions are especially  sensitive to the ways in which the mass of heavy quarks affects the formation of parton showers and can be used to  constrain the still not well understood dead cone effect in the QGP.   In the kinematic domain where parton mass plays the most important role, we predict a unique inversion of the mass hierarchy of jet quenching effects, with the modification of the 
momentum sharing distribution for prompt $b$-jets being the largest. This work opens a new direction of research on heavy flavor jet substructure in ultrarelativistic nuclear collisions and can be extended to different energy correlators in jets~\cite{Larkoski:2015kga}. It is already  useful in guiding the next generation of jet measurements in heavy ion reactions at RHIC and LHC, with experimental results on $b$-jet and $c$-jet substructure modification expected very soon.  

This work was supported by the U.S. Department of Energy under Contract No. DE-AC52-06NA25396, its Energy Early Career Program, and the Los Alamos National Laboratory LDRD program.

\vspace{-0.5cm}

\bibliographystyle{h-physrev}
\bibliography{mybib}

\end{document}